\documentclass[10pt,aps,pre,twocolumn,superscriptaddress]{revtex4-2}
\usepackage[linkcolor = blue, citecolor = red, urlcolor = blue, colorlinks = true]{hyperref}
\usepackage[usenames,dvipsnames,x11names]{xcolor}
\usepackage[normalem]{ulem}
\usepackage{graphicx}
\usepackage{listings}
\usepackage{stmaryrd}
\usepackage{amssymb}
\usepackage{amsmath}
\usepackage{gensymb}
\usepackage{float}
\usepackage{bbold}
\usepackage{bm}

\definecolor{crimson}{rgb}{0.75686,0,0.262745}
\definecolor{saphire}{rgb}{0.0,0.196,0.372549}
\definecolor{plum}{rgb}{0.50588,0.007843,0.3843137}

\begin{document}

\title{Thick liquid crystalline cholesteric shells}

\author{Arda Bulut}
\thanks{These authors contributed equally.}
 \affiliation{Physics Department, College of Sciences, Ko{\c c} University, Rumelifeneri Yolu 34450 Sar\i{}yer, Istanbul,  T{\" u}rkiye}
 
\author{Yusuf Sar\i{}yar}
\thanks{These authors contributed equally.}
 \affiliation{Physics Department, College of Sciences, Ko{\c c} University, Rumelifeneri Yolu 34450 Sar\i{}yer, Istanbul,  T{\" u}rkiye}

\author{Giuseppe Negro}
 \affiliation{SUPA, School of Physics and Astronomy, University of Edinburgh, Peter Guthrie Tait Road, Edinburgh, EH9 3FD, UK}
 
\author{Livio Nicola Carenza}
\email{lcarenza@ku.edu.tr}
 \affiliation{Physics Department, College of Sciences, Ko{\c c} University, Rumelifeneri Yolu 34450 Sar\i{}yer, Istanbul,  T{\" u}rkiye}


\begin{abstract}

We numerically investigate the phase behavior of thick shells of cholesteric liquid crystals with tangential anchoring at the shell boundary. 
For achiral liquid crystal, we demonstrate a thickness-dependent transition from a configuration featuring four disclination line connecting the inner and outer surfaces to a state free of defect in the bulk, where each surface is topologically isolated and features two boojums. 
Incorporating chirality stabilizes novel defect arrangements, including a mixed state combining boojums and disclination lines and blue phases at high chirality and we demonstrate that shell thickness strongly modulates these transitions.
Finally, we exploit the metastability features of the observed phases to obtain an elastically induced rearrangement of the shell surfaces during a cholesteric hysteresis cycle, stabilizing an alternative configuration that minimizes the free energy at low chirality. 
Our work paves the way for exploring dynamic behaviors under external fields, mixed anchoring conditions, or active flows.  

\end{abstract}

\maketitle
Liquid crystals (LCs) are a fascinating phase of matter of enduring interest in physics, serving as a versatile platform for studying and exploiting topological phases~\cite{de1993physics,Williams_1986,LAVRENTOVICH01011998,Lopez2011,PhysRevX.7.011006, Thot2002,Salinger2022}. Theoretically, LCs offer a unique opportunity to investigate topological phenomena in experimentally accessible systems, where the interplay between molecular chirality, confinement, and defect structures gives rise to complex and often surprising behaviors. From an applied perspective, LCs exhibit unique optical properties arising from their striking interaction with electromagnetic fields that have positioned LCs at the forefront of applications in nanotechnology, from photonic crystals~\cite{Laventrovich2011photonic,Money2023,Ma2022}, display devices and laser media~\cite{Humar2010,bluephaselasers}, to smart nano-materials~\cite{bluephaselasers,skyrmionssmalyukh,hopfionsmalyukh,Emre2016}, and sensig devices~\cite{Kurt2022}. This diverse behavior has motivated a wealth of research, highlighting the tunability of liquid crystals as a platform for creating novel materials with engineered optical properties.

Confinement plays a crucial role in shaping the optical properties of LC~\cite{Kamien2002,Nieves2007,Lopez2011,Napoli2012,Napoli2021,Ishii2020,copar2020,Carenza2024PRR}. By restricting LCs within specific geometries, the topological characteristics of the system--and thus its defect structures--can be precisely controlled. This is particularly impactful because the characteristic size of LC defects and distortions often match the wavelength of visible light, allowing LCs to act as exceptional anisotropic media with remarkable optical properties. A compelling demonstration of the interplay between topology and geometry is provided by liquid crystalline spherical droplets and shells with tangential anchoring. Both conform to the Poincaré–Hopf~\cite{Poincare,Hopf,Kamien2002} theorem, which dictates that the total topological charge on any spherical surface must equal the Euler characteristic of the sphere~\cite{Urbanski_2017,Laventrovich2010,Kamien2002,Mermin1979}, i.e. $+2$. In LC droplets, this constraint is satisfied by the formation of boojums, namely $+1$ defects positioned at the poles. In contrast, thin spherical shells accommodate this requirement through a configuration of four $+1/2$ defects arranged at the vertices of a regular tetrahedron~\cite{Lubensky1992,Leon2011}.

Chirality introduces a further feature to the study of liquid crystals (LCs), playing a central role in stabilizing topological phases~\cite{Kurik1981,Bezić01041992,Xu1997,Sec2012,Orlova2015,Zhou2016,tran2017,Bentancur2023,Kocaman2024,Pollard2018}. Blue phases (BPs) are a remarkable example~\cite{Wright1989,Alexander2006,Alexander_2008,Henrich2010,Henrich2011,kentbp3,Tanay2020,Musevich2022,Money2023}, characterized by networks of skyrmions that arise from topological frustration as the system accommodates helical twisting in multiple spatial directions, by patching together double-twist cylinders.
When confined, a cholesteric liquid crystal (CLC) must balance its intrinsic helical twist with the constraints imposed by finite geometry and its topology. This competition between topological and geometric frustration gives rise to novel phases, enabling precise tuning of LC properties by adjusting the geometry and size of the confining domain.  
This intricate interplay has been the focus of various studies. For example, de Pablo and collaborators investigated confined CLC droplets and blue phases~\cite{Gonzalez2015}, revealing that confinement not only broadens the typically narrow stability range of BPs but also alters their topological and optical characteristics. Moreover, some of the authors of the present study demonstrated numerically that CLCs confined to the surface of an infinitesimally thin shell exhibit fascinating phases that are robust regardless of the anchoring conditions on the sphere~\cite{carenza2022prl,Negrosoftmatter2023}. In such configurations, BPs appear as arrays of 2D skyrmions distributed across the spherical surface, forming skyrmionic quasi-crystals and amorphous patterns, further expanding the range of phases accessible to confined CLCs.

A particularly interesting setup is the one of \emph{thick} spherical shells of LC with degenerate tangential anchoring. 
This approach enables the systematic modulation of the competing effects of chirality-induced topological frustration and geometrical confinement, offering a flexible platform to explore the resulting phases and defect structures. Experimental studies on thick shells of chiral liquid crystals (CLC), such as those in Ref.~\cite{Darmon2016,Darmon2016PNAS}, have revealed a variety of topological defect arrangements. These include configurations beyond the classic bipolar shell and the four $+1/2$ defect arrangements, such as the stabilization of a Frank-Pryce-like structure featuring a single $+2$ defect, as well as intermediate defect configurations.

Inspired by prior studies on confined cholesteric liquid crystals, in this article we numerically investigate the interplay between topological and geometrical confinement in thick shells of CLCs, to build a phase diagram of their topological features at varying the shell thickness and LC chirality. To this end, we use lattice Boltzmann simulations to study the phase behavior of CLCs confined in a thick soft spherical shell modeled as dynamical phase fields, accounting for the intrinsic dynamics of the shell boundary mediated by surface tension. 


Our findings reveal how the constraints imposed by the Gauss-Bonnet theorem shape the defect configurations in shells of varying thickness. This requirement, which ensures a total topological charge of $+2$ on any spherical surface, governs the structures that can emerge under different geometric and chiral conditions. For sufficiently thick shells, we identify distinct phases depending on the chirality, including the bipolar structure, $+1/2$ disclinations connecting the inner and outer surfaces, and an intermediate phase that combines features of both. 
Interestingly, the transition from the boojum configuration to the four disclination lines can be induced either by reducing the shell thickness or increasing the chirality. In the latter case, the transition exhibits a metastable region occupied by intermediate states and pronounced hysteresis effects. For thin shells, the most stable configuration is characterized by four disclination lines, with a direct transition to BP.


\section{The model}
\label{sec:model}
To model thick cholesteric shells we consider a multiphase field approach~\cite{negro2024natcom,negro2023science} and introduce two scalar phase fields $\phi_i({\bm r}, t)$, with $i=1,2$, and the nematic $Q-$tensor ${\bm Q}({\bm r},t)$, whose principle eigenvector $\bm{n}$ --the so-called \emph{director field}-- defines the local direction of alignment of the liquid crystal (LC).  Finally, we consider the incompressible velocity field $\bm{v}({\bm r},t)$.
The equilibrium of the system is defined by the  free energy
\begin{equation}
    \mathcal{F} = \int \ d\bm{r} \ \left( f^{\phi} +  f^{\rm LC} + f^{\rm anch} \right) \; , 
    \label{eqn:1}
\end{equation}
where $f^{\phi}$ is the concentration free energy density, $f^{\rm LC}$ the liquid crystal free energy accounting for the isotropic-nematic transition and elastic properties of the LC, and $f^{\rm anch}$ defines the anchoring properties of the LC at the shell interface. In particular, 
\begin{equation}
   f^{\phi} = \sum_{i=1}^{2} \frac{a}{4}\phi_i^2(\phi_i-\phi_0)^2 + \frac{k_{\phi}}{2}\sum_{i=1}^{2}(\nabla\phi_i)^2  + \epsilon \phi_{1}^2\phi_{2}^2 \; .
   \label{eqn:2}
\end{equation}
For $a,k_{\phi}>0$, $f^{\phi}$ describes phase separating fields with two minima at $\phi_i=\phi_0$ and $\phi_i=0$ that, at equilibrium, arrange into spherical droplets with surface tension $\sigma=\sqrt{8ak_{\phi}/9}$, and interface thickness $\xi_{\phi} = \sqrt{2k_{\phi}/a}$. The interaction term proportional to $\epsilon$ gives rise to a repulsive contribution ($\epsilon>0$) that allows for the stabilization of the emulsion.
The LC free energy density $f^{\rm LC } =  f^{\rm bulk } + f^{\rm el }$, where
\begin{gather}
\begin{split}
f^{\rm bulk } = A_0 \left[ \dfrac{1}{2} \left( 1-\frac{\chi(\bar\phi)}{3} \right) {\rm Tr} {\bm Q}^2  -\frac{\chi(\bar\phi)}{3} {\rm Tr} {\bm Q}^3 \right. \\ \left. +\frac{\chi(\bar\phi)}{4} \left( {\rm Tr} {\bm Q}^2 \right)^2 \right] \; ,
\end{split}  \label{eqn:3}  \\
f^{\rm el} = f^{\rm sb} + f^{\rm tw} = \frac{L}{2}\left[ (\nabla \cdot \bm{Q})^2 + (\nabla \times \bm{Q} + 2q_0 \bm{Q} )^2 \right]\; \label{eqn:4} . 
\end{gather}

In Eq.~\eqref{eqn:3}, the bulk constant $A_0>0$, and $\chi(\bar\phi)$ is a temperature-like parameter that drives the isotropic-nematic transition that occurs for $\chi(\bar\phi)>\chi_{cr}=2.7$~\cite{de1993physics}, where we have denoted with $\bar\phi=\phi_1 + \phi_2$ the total concentration field. We choose $\chi=\chi_0+\chi_s\bar\phi$, with  $\chi_0>2.7$ and $\chi_s<0$ to confine the LC within the layer where $\bar\phi=0$. The elastic free energy density in Eq.~\eqref{eqn:4} proportional to the elastic constant $L$ captures the energy cost of elastic deformations in the single elastic constant approximation~\cite{de1993physics}. It has been split into a splay-bend contribution $f^{\rm sb}$ and a twist contribution $f^{\rm tw}$, which stabilizes helical structures when the chiral wavenumber $q_0$ is non-zero. For $q_{0}>0$, the equilibrium configuration in unconfined geometries features a right-handed helix with pitch $p_0=2\pi/q_0$.
Finally, the anchoring contribution to the free-energy
\begin{equation}
     f^{\rm anch } = W \sum_{i=1}^2 \left[\partial_{\alpha}\phi_i Q_{\alpha\beta} \partial_{\beta}\phi_i\right]. 
     \label{eqn:5}
\end{equation}
defines the orientation of the LC at the shell's interface. We choose $W>0$ to impose tangential anchoring to the surface.

The dynamics of the fields is governed by the following set of coupled PDEs,
\begin{eqnarray}
&D_t \phi_i = M \nabla^2 \mu_i \; , \label{eqn:6} \\
&D_t{\bm Q}= {\bm S}({\bm W},{\bm Q}) + \gamma^{-1} {\bm H} \; , \label{eqn:7} \\
&\rho D_t {\bf v} = \nabla\cdot ({\bm \sigma}^{\rm hydro}+{\bm \sigma}^{\phi}+{\bm \sigma}^{\rm LC}) \; , \label{eqn:8} 
\end{eqnarray}
where the operator $D_t= \partial_t + \bm{v}\cdot \nabla$ denotes the material derivative. Eq.~\eqref{eqn:6} is a Cahn-Hilliard equation for the conserved concentration field $\phi_i$, with $M$ the mobility parameter, and $\mu_i = \delta \mathcal{F}/\delta \phi_i$, for $i=1,2$.
Eq.~\eqref{eqn:7} is the Beris-Edwards equation ruling the dynamics of the $Q-$tensor. The operator $\bm{S}(\bm{W},\bm{Q})$ denotes the co-rotational derivative and defines the dynamical response of the LC to straining and shearing. Its explicit expression depends on both the velocity gradient $\bm{W} = \nabla \bm{v} $ and the $Q-$tensor configuration (see Eq.~\ref{eqnA1} for the explicit expression). 
The coefficient $\gamma$ is the rotational viscosity measuring the relative importance of advection with respect to relaxation, and the molecular field ${\bm H}=-\frac{\delta {\mathcal F}}{\delta {\bm Q}}+({\bm I}/3){\rm Tr}\frac{\delta {\mathcal F}}{\delta {\bm Q}}$.
Finally, Eq.~\eqref{eqn:8} is the Navier-Stokes equation for the incompressible velocity field ($\nabla \cdot \bm{v} = 0$) with constant density $\rho$. Here, the stress tensor has been divided in: \emph{(i)} a hydrodynamic contribution ${\bm \sigma}^{\rm hydro}= - P \bm{I} + \eta \nabla \bm{v}$ accounting for the hydrodynamic pressure $P$ ensuring incompressibility, and viscous effects, proportional to the viscosity $\eta$; \emph{(ii)} an interface contribution ${\bm \sigma}^{\phi}$, and \emph{(iii)} a LC contribution ${\bm \sigma}^{\rm LC}$ accounting for elastic and flow-aligning effects and whose explicit expressions are reported in Appendix~\ref{sec:app1b}.

\subsection*{Dimensionless numbers}
Thick LC shells can be obtained by encapsulating an isotropic droplet of the phase field $\phi_1$ with radius $R_1$ ($\phi_1=\phi_0$ for $r<R_1$) into an inverted larger droplet of the phase field $\phi_2$ with radius $R_2>R_1$ ($\phi_2=\phi_0$ for $r>R_2$).
This effectively realizes a shell of the total concentration field characterized by $\bar{\phi}=0$ where the LC is in the nematic phases, for $R_1<r<R_2$. 

In the following we will fix the radius of the external droplet $R_2$ and vary the radius $R_1$ of the internal core. A suitable dimensionless number to characterize the shell thickness is therefore
\begin{equation}
    r = \dfrac{R_2-R_1}{R_2} \; ,
    \label{eqn:9}
\end{equation}
that ranges between $0$ and $1$.
For the limit case of a thin shell $R_1 \approx R_2$ and $r \approx 0$; conversely for the case of a full droplet, in absence of an internal core, $R_1=0$, and $r=1$.

The cholesteric power of the LC will be expressed in terms of the reduced chirality
\begin{eqnarray}
    \kappa = \sqrt{\dfrac{108 q_0^2 L}{A_0 \chi_0}} \; ,
    \label{eqn:10}
\end{eqnarray}
proportional to the ratio $\ell_{\rm c}/p_0$, where the coherence length $\ell_{\rm c} = \sqrt{L/A_0}$ is  a measure of how rapidly nematic order decays in proximity of a topological defect.

Refer to Appendix~\ref{sec:app2a}, and~\ref{sec:app2b} for details on the simulation protocol and parameters.

\begin{figure}[htbp]
\centering
\includegraphics[width=1\columnwidth]{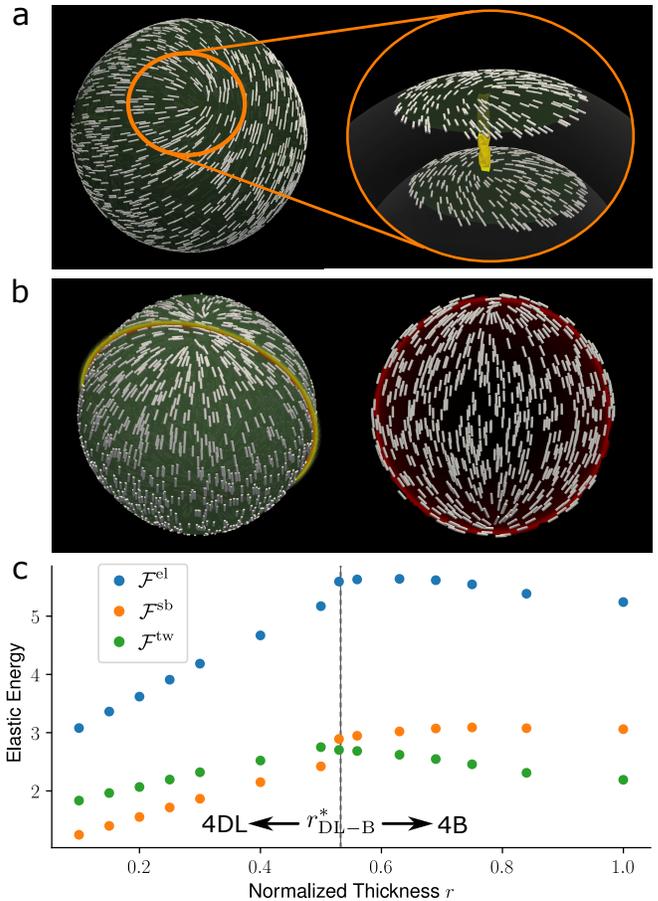}
\caption{\textbf{Nematic shells.} (a) Thin LC shells are characterized by four $+1/2$ topological defects. The defects on the inner and outer surface of the shell are terminal points of an open disclination line (yellow line in the zoom). (b) Full nematic LC droplets arrange into a bipolar configuration, featuring two antipodal \emph{boojums}. These defective structures consist of a $+1$ surface defect that escapes in the third dimension in the droplet bulk, leaving the droplet's interior defect-free. Right panel shows the nematic arrangement in the droplet's cross-section along the plane highlighted in yellow on the left. (c) Elastic energy $\mathcal{F}^{\rm el}$ at varying the shell thickness $r$. Dashed grey line marks the transition point between $\rm{4DL}$ and $4B$ configurations at $r^{*}_{\rm DL-B}=0.53$.} 
\label{fig1}
\end{figure} 

\begin{figure*}[htbp]
\centering
\includegraphics[width=0.7\textwidth]{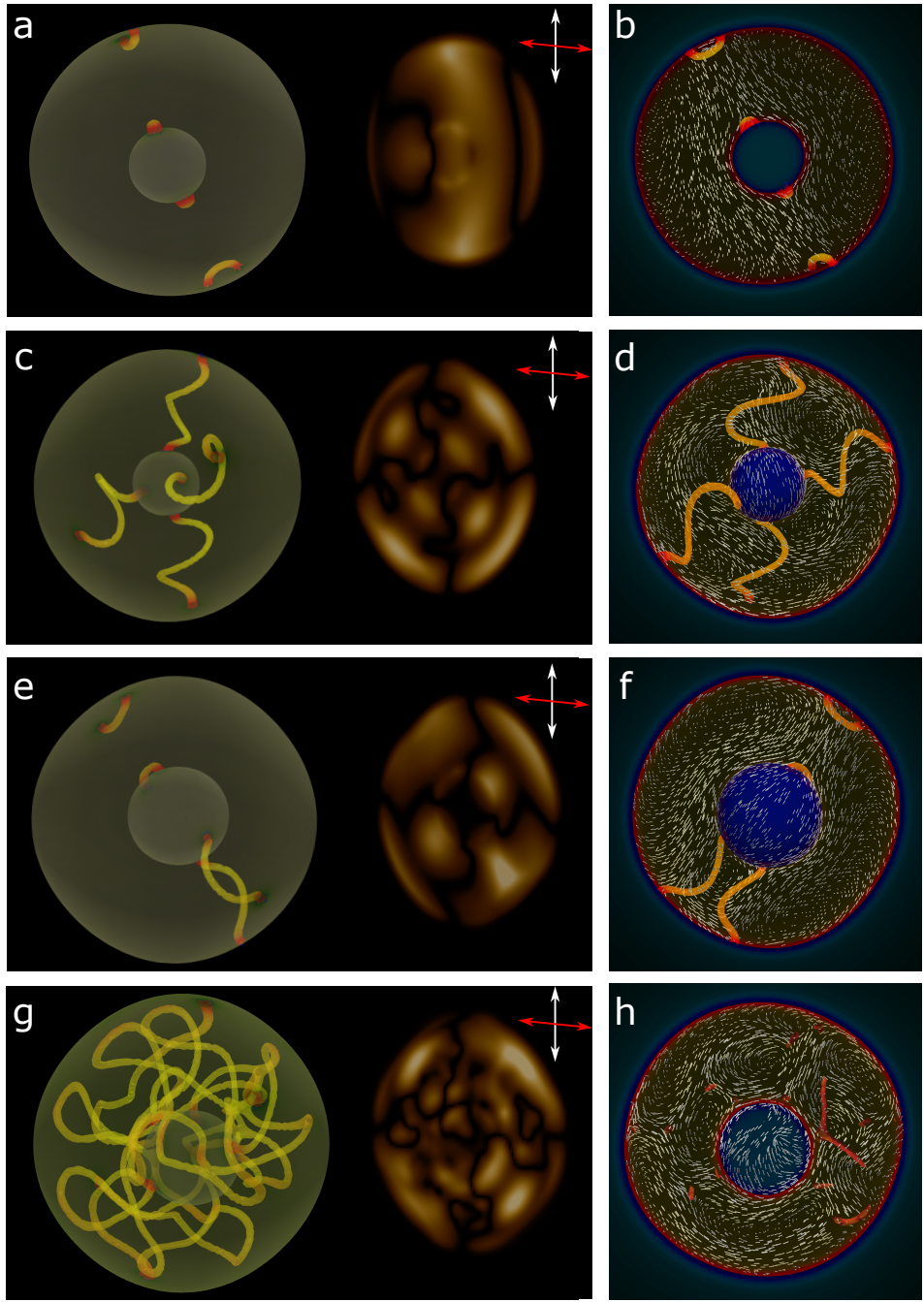}
\caption{\textbf{CLC shells.} Typical defect configurations, cross-polarizer textures, and bulk arrangement of director field for: (a-b) $\rm{4B}$ state ($r=0.69; \kappa=0.2$); (c-d) $\rm{4DL}$ state ($r=0.72; \kappa=0.6$); (e-f) $\rm{2B-2DL}$ state ($r=0.6; \kappa=0.64$); and (g-h) ${\rm BP}$ state ($r=0.62; \kappa=0.84$). The red-white arrows at the top right corner of the panels on the left define the orientation of the cross-polarizer.}
\label{fig2}
\end{figure*} 

\section{Non-chiral nematic shells}
\label{sec:non-chiral}
We begin our discussion by analyzing the topological properties and defect configurations of nematic shells with variable thickness $r$ in the absence of chirality ($ \kappa = 0$).

In the limiting case of an infinitesimally thin shell ($r = 0$), the liquid crystal (LC) behaves as a two-dimensional system, and the equilibrium configuration is characterized by four defects of charge $s=1/2$ arranged in the so-called \emph{baseball ball} configuration, summing to a total topological charge of $ +2 $, as required by the Poincaré–Hopf theorem (see Fig.~\ref{fig1}a). While the topological constraint allows for any defect configuration with a total charge equal to the Euler characteristic of the sphere \( \chi_E = 2 \), the equilibrium state minimizes the free energy, as described by Eq.~\eqref{eqn:1}, for the given geometry. For instance, the topological requirement could also be satisfied by two integer defects of charge \( s = +1 \). However, such a configuration is energetically unfavorable because the energy of a two-dimensional defect scales with the square of its charge, \( \mathcal{F}^{\rm def} \sim s^2 \). Consequently, while topologically equivalent, a configuration with four half-integer defects is energetically preferred, as it reduces the defect energy by half compared to the two-integer-defect arrangement.
 \begin{figure*}[htbp]
\centering
\includegraphics[width=0.85\textwidth]{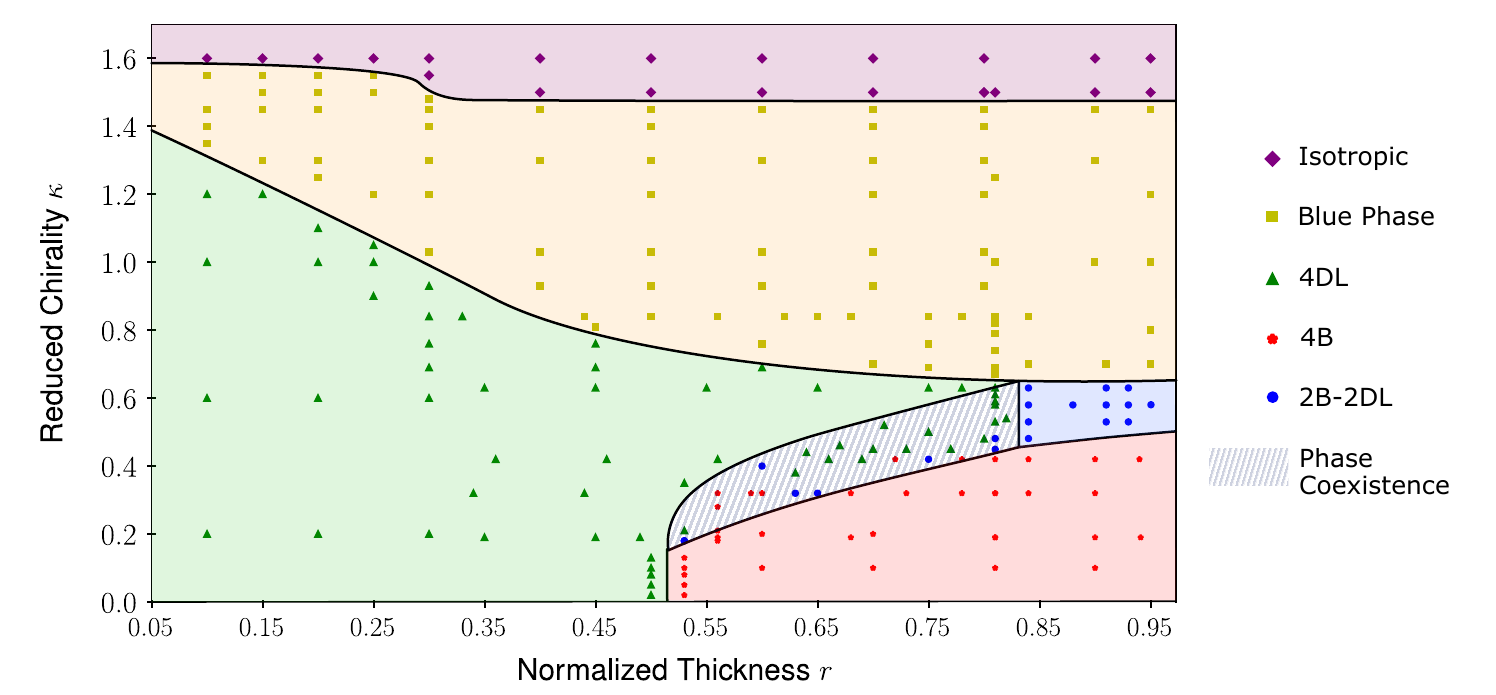}
\caption{\textbf{Phase diagram.} Phase diagram at varying shell thickness $r$ and chirality $\kappa$.} 
\label{fig3}
\end{figure*} 
When the shell thickness $r$ increases to finite values, the topology of the system changes. Now, the topological constraint must be satisfied on both the internal and external surfaces of the LC shell. For thin shells, the  baseball ball pattern of four defects is reproduced across concentric spherical surfaces, from the outer to the inner shell surface. This leads to the stabilization of four disclination lines connecting the $ +1/2 $ defects on the outer and inner surfaces of the shell, as illustrated in the magnified view in panel~a of Fig.~\ref{fig1}.

For the limit case of a full LC droplet ($r=1$), the topological requirement of a total charge of $+2$ on the droplet surface remains unchanged. However, in this case, the energetically favored configuration must minimize the elastic energy of the whole droplet, not just the surface. This leads to a drastic change in the defect arrangement. 
The four disclination lines give way to the \emph{bipolar} configuration, {\rm i.e.} two antipodal \emph{boojums} on the droplet surface. These singular structures consist of a surface $+1$ defect that escapes into the third dimension within the droplet's bulk, leaving it free of disclination lines (see Fig.~\ref{fig1}b).
In this regard it is important to remember that  the radial symmetry of $+1$ defects allows the director to rotate away from the plane of the defect, so that the singular behavior in the bulk \emph{softly and suddenly vanishes away}~\cite{Mermin1979,Mermin_1990}. This is in stark contrast to semi-integer defects, where a continuous out-of-plane rotation of the director around the singularity would introduce a discontinuity along a ray extending from the singularity—commonly referred to as \emph{Volterra's cut}.  

Interestingly, we observe a configuration analogous to the bipolar arrangement for very thick shells ($r \lesssim 1$). In these cases, the topological requirement on the inner shell surface is satisfied by the emergence of two additional boojums, while the nematic pattern in the bulk remains devoid of disclination lines (see Fig.~\ref{fig2}a,b for reference).  

Having described the two limiting cases, we now address the transition between these states as the shell thickness $r$ varies. Specifically, we find that the system undergoes a transition from a configuration with four disclination lines connecting the inner and outer surfaces ($\rm{4DL}$) to one characterized by four boojums ($\rm{4B}$), where the defect lines vanish entirely from the bulk. This transition occurs at a critical shell thickness of $r^{*}_{\rm DL-B}=0.53$. 
We rationalize this transition as a balance between the elastic energy contributions from the bulk and the surfaces. For configurations with disclination lines, the $+1/2$ surface defects are energetically favored over integer defects due to their lower surface energy. However, the energy of the disclination lines scales linearly with their length, making them increasingly costly as the shell thickness grows.
In contrast, boojums, despite their higher surface energy due to the $+1$ charge, completely eliminate disclination lines from the bulk. This results in a weakly deformed nematic director field within the bulk, significantly reducing the bulk elastic energy for $r > r^{*}_{\rm DL-B}$.  
This is evidenced by the behavior of the splay-bend contribution to the elastic energy, which stabilizes at a constant value beyond the transition (see Fig.~\ref{fig1}c). This is primarily attributed to localized splay deformations near the boojums. Meanwhile, the twist contribution to the elastic energy decreases as the shell thickness increases, further minimizing the total elastic energy.

\section{Cholesteric LC shells}
\label{sec:chiral}
We now turn our attention to cholesteric liquid crystalline shells. In the presence of chirality, we observe a wide range of configurations that depend on both the chirality of the liquid crystal and the shell thickness. We begin by describing their topological properties and then discuss their stability.  

\paragraph*{4B.} As outlined in the previous section, the topological constraint of a total $ +2 $ topological charge can be satisfied by the formation of two boojums on both the inner and outer surfaces of the shell (Fig.~\ref{fig2}a). When chirality is introduced, the internal arrangement of the liquid crystal adapts to incorporate a twisting deformation with a pitch determined by the equilibrium cholesteric wavenumber $q_0$ (Fig.~\ref{fig2}b). Under cross-polarized light oriented parallel to the axis defined by the four boojums, the transmitted light intensity is enhanced along this axis and displays a characteristic modulation in the transverse direction, reflecting the twisted liquid crystal pattern. 

\paragraph*{4DL.} In the presence of chirality, the ${\rm 4DL}$ structure retains its defining feature of four disclination lines connecting semi-integer defects on the inner and outer surfaces of the shell. Due to the imposed equilibrium twist, the disclination lines adopt a helical configuration (Fig.~\ref{fig2}c). In particular, at sufficiently high chirality, the liquid crystal in the bulk forms skyrmionic structures that emerge between each pair of disclination lines, as illustrated in the cross-section shown in panel~d of Fig.~\ref{fig2}. Cross-polarizer experiments reveal a distinctive quadrupolar pattern, with shadowed regions corresponding to the positions of the four disclinations and intensity modulations indicative of the internal twisted arrangement.
\begin{figure}[t!b]
\centering
\includegraphics[width=1\columnwidth]{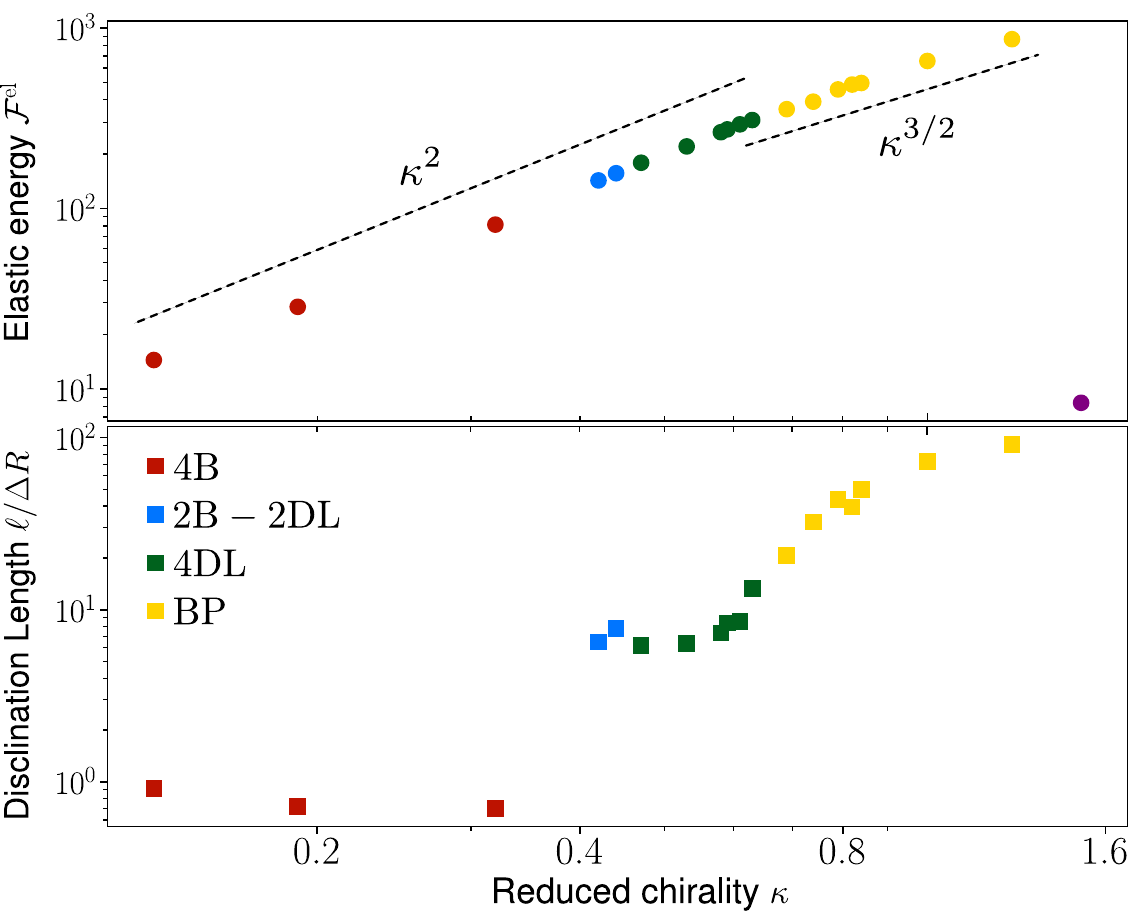}
\caption{\textbf{Characterization as a function of $\kappa$ for $r=0.81$.} The top panel presents the total elastic energy, $\mathcal{F}^{\rm el}$, plotted on a log-log scale against the reduced chirality $\kappa$. Dashed lines serve as visual guides, indicating slopes proportional to $\kappa^2$ (left) and $\kappa^{3/2}$ (right). The bottom panel displays the average measured length of the disclination lines. Data points in both panels are color-coded according to the phase diagram in Fig.~\ref{fig3}, as specified in the legend shown in the bottom panel.}
\label{fig4}
\end{figure} 
\paragraph*{2B-2DL.} Not encountered in absence of chirality, the ${\rm 2B-2DL}$ structure is a mixed structure resulting from the combination of the ${\rm 4B}$ and ${\rm 4DL}$ configuration. It features two boojums (one on the internal and one on the external surface) and two disclination lines connecting semi-integer surface defects on the internal and external surface (Fig.~\ref{fig2}e). The resulting configuration in the bulk is characterized by an escaped configuration in the region occupied by the two boojums and a skyrmionic structure separating the two disclination lines. The cross-polarizer textures reveal a strong luminosity in correspondence of the skyrmionic structure (Fig.~\ref{fig2}f). Notably, we observe that the inner core of the shells exhibiting a ${\rm 2B-2DL}$ structure is \emph{slightly} displaced from the center, shifting toward the shell surface along the direction of the disclination lines.

\paragraph*{BP.} The blue phase (BP) is a topological phase that emerges at sufficiently high chirality and is characterized by a three-dimensional network of disclination lines (Fig.\ref{fig2}g), which accommodate an array of double-twist cylinders in the director field (Fig.\ref{fig2}h). In our simulations, the disclination network does not exhibit periodicity, resulting in a structure that resembles an unconfined blue phase of type III, commonly referred to as \emph{blue fog} \cite{Wright1989,Henrich2011} due to its diffuse and irregular optical appearance under cross-polarizers.

\subsection*{Phase behavior of CLC shells}
We now analyze the stability of the four states discussed in the previous section as a function of shell thickness $r$ and reduced chirality $\kappa$. The complete phase diagram of the system is shown in Fig.~\ref{fig3}.

In Section~\ref{sec:non-chiral}, we described the transition from the ${\rm 4DL}$ configuration (green region Fig.~\ref{fig3}) to the ${\rm 4B}$ configuration (red region) as the shell thickness exceeds the critical value $r^{*}_{\rm DL-B}$. For low chirality ($\kappa^{*} \simeq 0.2$), this transition remains unchanged. This is because for chirality to affect the transition, the equilibrium pitch must be comparable to the shell thickness, i.e., $p_0 \approx R_2 r^{*}_{\rm DL-B}$, which corresponds to $\kappa = 0.22$, in agreement with our observations.


As chirality increases beyond $\kappa^{*}$, the stability range of the ${\rm 4DL}$ configuration extends to thicker shells. This occurs because the topological frustration introduced by chirality forces the liquid crystal in the bulk to a chiral arrangement, leading to helical disclination lines that accommodate skyrmionic structures (Fig.~\ref{fig2}c-d). 
Notably, the transition between ${\rm 4DL}$ and ${\rm 4B}$ regions in the phase diagram of Fig.~\ref{fig3} is now separated by a phase co-existence region (shaded in Fig.~\ref{fig3}), where ${\rm 4DL}$, and ${\rm 4B}$, and ${\rm 2B-2DL}$ configurations are observed. The ${\rm 2B-2DL}$ configuration stabilizes for thicker shells ($r \gtrsim 0.82$) and intermediate chirality values ($0.4 \lesssim \kappa \lesssim 0.6$) colored blue in Fig.~\ref{fig3}. 

\begin{figure*}[htbp]
\centering
\includegraphics[width=1\textwidth]{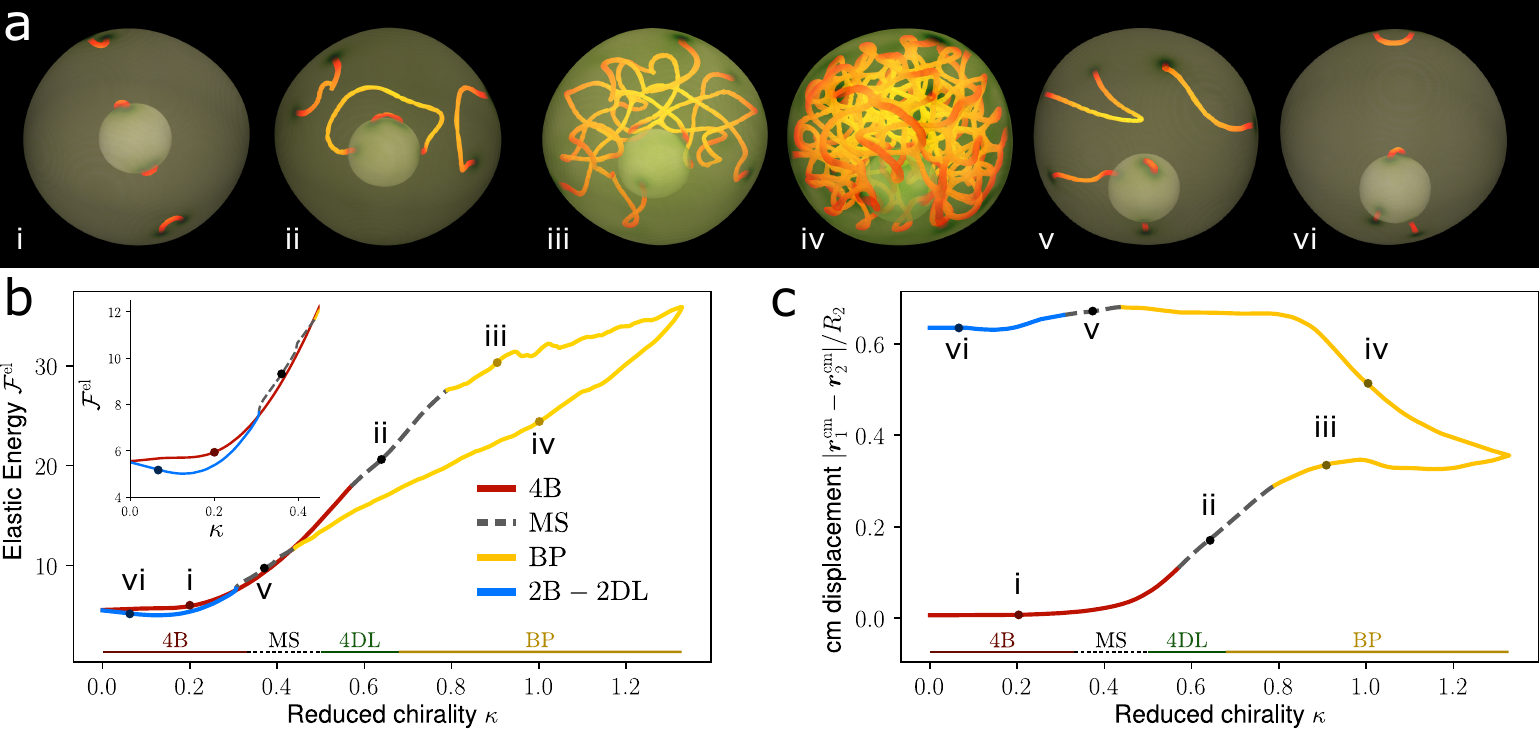}
\caption{\textbf{Hysteresis test at $r=0.66$.} (a) Gallery of configurations observed during the hysteresis test, with disclination lines highlighted in yellow. (b) Elastic free energy, $\mathcal{F}^{\rm el}$, as a function of reduced chirality, $\kappa$. The inset provides a closer view of the low-chirality region during the hysteresis test. (c) Relative displacement of the center of mass of the internal droplet, $\bm{r}^{\rm cm}_1$, with respect to the center of mass of the external droplet, $\bm{r}^{\rm cm}_2$, normalized by the radius of the external droplet, $R_2$. The curves in panels (b) and (c) are color-coded to correspond to the configurations described in the legend of panel (b). Colored dots in panels (b) and (c), labeled with Roman numerals, indicate the configurations shown in panel (a). The colored bars at the bottom of panels (b) and (c) denote the observed phases, as reported in the phase diagram of Fig.~\ref{fig3} for \(r=0.66\).} 
\label{fig5}
\end{figure*} 


For higher chirality values ($\kappa > \kappa_{\rm BP}^{*}=0.6$), BP states emerge for thick shells (yellow region in the phase diagram of Fig.~\ref{fig3}). This roughly corresponds to the transition points between a helical state and the BP in an unconfined geometry (not shown).  Interestingly, the frustration of the helical LC pattern induced by the shell thickness at small $r$, allows for the  stabilization of ${\rm 4DL}$ states even at large chirality. This effect is particularly evident for $r < r^{*}_{\rm DL-B}$, consistent with the requirement that the pitch $p_0$ of the blue phase’s double-twist cylinders matches the shell thickness. Finally, for $\kappa > \kappa_{\rm ISO}=1.5$, the BP structure transitions to the isotropic phase (purple region Fig.~\ref{fig3}). 

To better characterize the transitions between the various phases described above, Fig.~\ref{fig4} shows the elastic energy $\mathcal{F}^{\rm el}$ and the average length of the disclination line $\ell$ (normalized with respect to the shell's thickness $\Delta R$) as functions of chirality $\kappa$ for a shell with reduced thickness $r=0.81$. [Refer to Sect.~\ref{sec:app3a} for details on the numerical implementation of disclination line detection.] At low chirality, the shell adopts the ${\rm 4B}$ configuration, where disclination lines are effectively absent. In this regime, the elastic energy increases quadratically with chirality, following $\mathcal{F}^{\rm el} \sim \kappa^2$, as the liquid crystal develops a helical arrangement. 
As chirality increases, the system transitions to the region characterized by phase coexistence, where it alternates between the ${\rm 2B-2DL}$ and ${\rm 4DL}$ configurations. This transition is marked by a significant rise in the average length of the disclination lines. With further increases in chirality, the disclination lines lengthen as they develop helical structures that extend beyond the shell thickness $\Delta R$, while the elastic energy continues to scale as $\mathcal{F}^{\rm el} \sim \kappa^2$. 
When the system enters ${\rm BP}$, the length of the disclination lines $\ell$ continues to grow due to the formation of the network of double-twist cylinders and their shrinkage as $\kappa$ increases. Furthermore, the transition to the ${\rm BP}$ is signaled by a change in the scaling behavior of the elastic energy, which now follows $\mathcal{F}^{\rm el} \sim \kappa^{3/2}$. 
Finally, as chirality exceeds $\kappa_{\rm ISO}$, the system transitions to the isotropic phase, causing the elastic energy to drop.

\section{Hysteresis test}

To further investigate the nature of the transitions between configurations, we perform a hysteresis test by quasi-statically increasing the chirality of the CLC in a shell with reduced thickness \( r = 0.66 \). Starting from a chirality-free initial state in the \({\rm 4B}\) configuration (Fig.~\ref{fig5}a(i)), we tracked the changes in configuration by plotting the elastic free energy \(\mathcal{F}^{\rm el}\) as a function of reduced chirality in Fig.~\ref{fig5}b. Notably, the \({\rm 4B}\) configuration remained stable up to \(\kappa \simeq 0.57\), well beyond the critical value \(\kappa_{\rm MS-DL}(r = 0.66) \simeq 0.5\), indicating strong metastability within the phase coexistence region identified in the phase diagram.  
Beyond \(\kappa \simeq 0.57\), the system transitioned into spurious configurations, as shown in Fig.~\ref{fig5}a(ii). Here, the two boojums on the external surface split into pairs of semi-integer surface defects connected by disclination lines. On the inner shell, one boojum remained stable, while the other split into semi-integer defects connected by a disclination running through the droplet's bulk. This rearrangement significantly affected the internal shell surface, which shifted in the direction of lower disclination density. To quantify this effect, Fig.~\ref{fig5}c plots the relative displacement of the internal droplet's center of mass \(\bm{r}_1^{\rm cm}\) with respect to the external droplet's center of mass \(\bm{r}_2^{\rm cm}\). The internal droplet's asymmetry increased markedly with chirality and stabilized once the system entered the \({\rm BP}\) phase at $r\simeq0.75$.
In this phase, the displaced internal core favored the formation of a network of double-twist cylinders in the region vacated by the core (Fig.~\ref{fig5}a(iii)). Upon further increasing \(\kappa\), the center-of-mass displacement stabilized until we reversed the process by quasi-statically decreasing chirality. During this stage, the inner core was pushed rapidly towards the outer shell, as shown in Fig.~\ref{fig5}a(iv) and Fig.~\ref{fig5}c. This effect arose because the decreasing chirality enlarged the double-twist cylinders, generating elastic pressure in disclination-rich regions that progressively pushed the inner core outward. This process also caused the elastic free energy to decrease more rapidly than during the chirality-increasing phase (Fig.~\ref{fig5}b). 
Below \(\kappa \simeq 0.4\), the disclination network collapsed, leaving behind metastable spurious configurations (Fig.~\ref{fig5}a(v)), which eventually transitioned to the \({\rm 2B-2DL}\) state at \(\kappa \simeq 0.3\). However, unlike the thermalization experiments used to draw the phase diagram of Fig.~\ref{fig3} , which began from a high-temperature isotropic state, the internal core in this case remained off-center, almost contacting the external surface (Fig.~\ref{fig5}a(vi)). In this configuration, the disclination lines had minimal extension, enabling a defect-free bulk configuration in the limit of vanishing chirality. The energy of this state matched that of the \({\rm 4B}\) configuration (Fig.~\ref{fig5}a(i)), as shown in the inset of Fig.~\ref{fig5}b.

\section{Discussion and Conclusions}

In this study, we numerically investigated the interplay between topological and geometrical constraints in thick cholesteric liquid crystal (CLC) shells. Our numerical approach enables the simulation of the full nemato-hydrodynamic equations in a shell geometry of arbitrary thickness while accounting for the intrinsic dynamics of the shell boundary mediated by surface tension and anchoring couplings with the confined LC.  
For nematic LC shells without chirality, we demonstrate that increasing shell thickness drives a transition from a ${\rm 4DL}$ configuration, where the inner and outer shell surfaces are connected by four disclination lines, to a configuration in which the two shell surfaces are topologically isolated, each featuring two boojums (${\rm 4B}$). This transition arises from a balance between elastic bulk and surface energies. Thin shells favor the ${\rm 4DL}$ arrangement due to the reduced surface energy of $s = 1/2$ defects, whereas thicker shells minimize bulk elastic energy by adopting the ${\rm 4B}$ state, which eliminates disclination lines.  
Introducing chirality stabilizes new configurations. At low chirality, corresponding to a pitch larger than the shell thickness, the behavior remains analogous to nematic shells, with transitions dominated by shell thickness alone. At higher chirality, however, the system stabilizes novel states. Chiral frustration leads to helical distortions of the disclination lines and the emergence of skyrmionic structures. For intermediate chirality and thick shells, we observe a mixed state (${\rm 2B-2DL}$), which combines boojums and disclination lines, balancing elastic energy contributions from bulk and surface. At high chirality, the system transitions to the ${\rm BP}$ state, characterized by a disordered network of double-twist cylinders, resembling the unconfined blue fog phase. Interestingly, the transition to the ${\rm BP}$ state is strongly influenced by shell thickness, with thinner shells delaying the onset of this phase.  
Our simulations also reveal phase coexistence between ${\rm 4DL}$, ${\rm 4B}$, and ${\rm 2B-2DL}$ states, suggesting a rich energy landscape with multiple local minima. This underscores the importance of the system's history, initial conditions, and external perturbations in determining the equilibrium configuration.  

Our findings are qualitatively consistent with prior experimental investigations of CLC shells. In Refs.~\cite{Darmon2016,Darmon2016PNAS}, Darmon \emph{et al.} observed five distinct defect configurations, including three reported here: ${\rm 4B}$, ${\rm 4DL}$, and ${\rm 2B-2DL}$. The experiments also identified two additional states: one with two disclination lines connecting the inner and outer surfaces of the droplet, exhibiting $+1/2$ and $+3/2$ profiles, respectively, and another featuring a single defect on each surface with a surface topological charge of $+2$, connected by a Dirac string (also known as  Frank-Pryce structure~\cite{Lopez2011}). These configurations were not observed in our simulations, likely due to their metastability, as suggested by the low occurrence of the first state in experiments, or due to our choice of elastic constants and the single-elastic-constant approximation used in our numerical model.  

In conclusion, our results provide a comprehensive phase diagram for nematic and cholesteric LC shells, qualitatively consistent with prior experimental observations. This work bridges fundamental topological principles with the complex interplay of shell geometry and material chirality. The ability to systematically explore and stabilize distinct topological states paves the way for designing responsive and tunable LC-based materials.  
These findings form the basis for future investigations into the dynamic behavior of these systems under external stimuli, mixed boundary conditions, or in multi-core emulsions. Additionally, it would be intriguing to explore how activity and self-sustained flows~\cite{ruske2021,Carenza22065, carenza2020_physA, Head2024,Metselaar2019,negro2024natcom,Jiron2024,Kralj2024} influence the topological states observed at equilibrium.  

\begin{acknowledgments}
L.~N.~C. acknowledges the support of the Postdoctoral EMBO Fellowship ALTF 353-2023 and the TÜBİTAK 2232/B program (project no. 123C289). The authors thank Emre B{\" u}k{\" u}{\c s}o{\v g}lu and Ludwig Hoffmann for insightful discussions.
\end{acknowledgments}

\appendix

\section{Nemato-hydrodynamics}
\label{sec:app1}

\subsection{Corotational derivative} 
\label{sec:app1a}
The explicit expression of the corototational derivative $\bm{S}({\bm W},{\bm Q})$ appearing in Eq.~\eqref{eqn:7} is given by
\begin{multline}
\bm{S}({\bm W},{\bm Q})=(\xi{\bm D}+{\bm \Omega})({\bm Q}+{\bm I}/3)\\
+ (\xi{\bm D}-{\bm\Omega})({\bm Q}+{\bm I}/3) \\
-2\xi({\bm Q}+{\bm I}/3) {\rm Tr} ({\bm Q}{\bm W}). \; ,
\label{eqnA1}
\end{multline}
Here, ${\bm D}=({\bm W}+{\bm W}^T)/2$ and ${\bm\Omega}=({\bm W}-{\bm W}^T)/2$ are the symmetric and anti-symmetric part of the velocity gradient tensor $W_{\alpha\beta}=\partial_{\beta}v_{\alpha}$, respectively. The flow alignment parameter $\xi$ determines the aspect-ratio of the LC molecules and the dynamical response of the LC to an imposed shear flow. Here, we choose $\xi = 0.7$ to consider flow-aligning rod-like molecules.

\subsection{Stress tensor}
\label{sec:app1b}
The explicit expression of the interface contribution $\bm{\sigma}^{\phi}$ to the total stress is given by
\begin{equation}
\sigma^{\phi}_{\alpha\beta}=\sum_{i=1}^2\left[\left(f-\phi_i\frac{\delta{\cal F}}{\delta\phi_i}\right)\delta_{\alpha\beta}-\frac{\partial f}{\partial(\partial_{\beta}\phi_i)}\partial_{\alpha}\phi_i\right] \; ,
\label{eqnA2}
\end{equation}
while the explicit expression of the LC contribution is 
\begin{eqnarray}
\sigma_{\alpha\beta}^{LC}=&&-\xi H_{\alpha\gamma}(Q_{\gamma\beta}+\frac{1}{3}\delta_{\gamma\beta})-\xi(Q_{\alpha\gamma}+\frac{1}{3}\delta_{\alpha\gamma})H_{\gamma\beta}\nonumber\\
&&+2\xi(Q_{\alpha\beta}-\frac{1}{3}\delta_{\alpha\beta})Q_{\gamma\mu}H_{\gamma\mu}+Q_{\alpha\gamma}H_{\gamma\beta}-H_{\alpha\gamma}Q_{\gamma\beta}\nonumber\\
&&-\partial_{\alpha}Q_{\gamma\mu}\frac{\partial f}{\partial(\partial_{\beta}Q_{\gamma\mu})}.
\end{eqnarray}

\section{Numerical experiments}
\subsection{Simulation protocol}
\label{sec:app2a}
We integrate the dynamics of the hydrodynamic fields in Eq.~\eqref{eqn:6}-\eqref{eqn:8} in a cubic grid of size $\mathcal{L}=128$ with a predictor-corrector hybrid lattice Boltzmann approach~\cite{succi2018,carenzaepje} consisting of solving Eq.~\eqref{eqn:6},\eqref{eqn:7}  with a finite-difference algorithm implementing first-order upwind scheme and fourth-order accurate stencils for space derivatives, and the Navier-Stokes equation through a predictor-corrector LB scheme on a $D3Q15$ lattice. For technical details on the lattice Boltzmann method here implemented refer to References~\cite{carenzaepje,Negro2024}.

For the quenching experiments from high-temperature configurations used to produce the phase diagram in Fig.~\ref{fig3}, the system is initialized in the following manner. The concentration field $\phi_1$ is set as a spherical droplet with radius $R_1$, where $\phi_1=\phi_0$ inside the droplet and $\phi_1=0$ outside it. Similarly, the concentration field $\phi_2$ is initialized as a spherical droplet with radius $R_2>R_1$, where $\phi_2=0$ inside and $\phi_2=0$ outside. This configuration produces a global concentration field $\bar{\phi}=\phi_1+\phi_2$ that is zero within a spherical shell between the internal radius $R_1$ and the external radius $R_2$, and reaches the value $\phi_0$ outside this shell and within the internal inclusion.
The $Q-$ tensor is initialized in an isotropic state outside the shell, while inside the shell it is initialized with a small amplitude and a random orientation.
The velocity field is initialized to $0$.

For the hysteresis test presented in Fig.~\ref{fig5}, we started from a ${\rm 4B}$ configuration obtained in absence of chirality and reduced thickness $r=0.66$, according to the previous protocol. Then, we have steadily increased the chirality at a rate of $\Delta q_0 / \Delta t = 3.5 \times 10^{-6} {\rm [L]^{-1}/iteration}$, corresponding to a variation of the reduced chirality of $\Delta \kappa / \Delta t = 2 \times 10^{-6} {\rm (iteration)}^{-1}$.

\subsection{Simulation parameters}
\label{sec:app2b}
In this study we have varied two control parameters: \emph{(i)} the radius $R_1$ of the internal droplet was varied between $0$ and $29$, corresponding to a variation of the normalized radius $r$ from $1$ to $0.1$; \emph{(ii)} the chiral power $q_0$ was varied between $0$ and $0.54$, corresponding to a variation of the reduced chirality between $0$ and $1.6$.
All other parameters were kept fixed. In particular: $a=0.091, k_{\phi}=0.13, \epsilon=0.01, M=0.35$; $A_0=0.13, \chi_0=2.85, \chi_s=0.25, L=0.03, W=0.05. \xi=0.7, \gamma=1.0$. Finally the viscosity $\eta=3.3$.

\section{Measurements}

\subsection{Analysis of the center of mass}
\label{sec:app3a}
The center of mass of each droplet has been computed as
$$
\bm{r}_i^{\rm cm} = \int \text{d}\mathbf{r} \, \vartheta_H \left(\phi_i - \dfrac{\phi_0}{2} \right)   \mathbf{r}  \; , 
$$ 
where $\vartheta_H(x)$ is the Heavside theta function that is $1$ for $x \geq 0$ and $0$ otherwise.

\subsection{Disclination line identification}
\label{sec:app3b}
To measure the length of the disclination lines we have used the disclination density tensor, proposed by Schimmings and Vi\~nals \cite{schimming2022}. The tensor is constructed from derivatives of the Q-tensor
\begin{align}    D_{ij}=\epsilon_{i\mu\nu}\epsilon_{jlk}\partial_l Q_{\mu\alpha}\partial_k Q_{\nu\alpha}
\end{align}
where $i,j,k,\alpha,\mu,\nu$ are tensor indices with applied summation convention. 
The tensor can be decomposed in the direct product of the local line tangent $\mathbf{T}$ and the rotation vector $\mathbf{\Omega}$, so that
\begin{align}
    \label{eq:Dtensordecomposition}
    D_{ij} = s(\mathbf{r})\mathbf{\Omega}_{i}\mathbf{T}_j,
\end{align}
where $s(\mathbf{r})$ is a positive scalar field that is maximum at the disclination core. 
The contour lengths are found by grouping the disclination cells $s(\mathbf{r}) \geq0.033$ into an ordered sequence of points that form, followed by a line interpolation.

\bibliography{biblio.bib}

\end{document}